\documentclass[prb,twocolumn,aps,showpacs]{revtex4}
\usepackage{graphicx,color}
\usepackage{amsmath}
\usepackage{amsfonts, amssymb, amstext, amsxtra,times}
\usepackage[dvips]{hyperref}

\DeclareMathOperator{\Tr}{tr}

\DeclareMathOperator{\iim}{Im}

\newcommand{\al}{\alpha}
\newcommand{\ep}{\epsilon}
\newcommand{\be}{\begin{equation}}
\newcommand{\ee}{\end{equation}}

\begin{document}

\title{Effective Equilibrium Description of Nonequilibrium Quantum Transport I:\\Fundamentals and Methodology}

\author{Prasenjit Dutt$^1$, Jens Koch$^1$, J.\ E. Han$^2$, Karyn Le Hur$^1$}
\affiliation{$^1$ Departments of Physics and Applied Physics, Yale University, New Haven, CT 06520, USA}
\affiliation{$^2$ Department of Physics, State University of New York at Buffalo, Buffalo, NY 14260, USA}
 
\date{\today} 
\begin{abstract}
The theoretical description of strongly correlated quantum systems out of equilibrium presents several challenges and a number of open questions persist. In this paper we focus on nonlinear electronic transport through a quantum dot maintained at finite bias using a concept introduced by Hershfield [Phys.\ Rev.\ Lett.\ \textbf{70}, 2134 (1993)] whereby one can express such nonequilibrium quantum impurity models in terms of the system's Lippmann-Schwinger operators. These scattering operators allow one to reformulate the nonequilibrium problem as an effective equilibrium problem associated with a modified Hamiltonian, thus facilitating the implementation of equilibrium many-body techniques. We provide an alternative derivation of the effective Hamiltonian of Hershfield using the concept of an ``open system".
\ Furthermore, we demonstrate the equivalence between observables computed using the Schwinger-Keldysh framework and the effective equilibrium approach.
\ For the study of transport, the non-equilibrium spectral function of the dot is identified as the quantity of principal interest and we derive general expressions for the current (the Meir-Wingreen formula) and the charge occupation of the dot. We introduce a finite temperature formalism which is used as a tool for computing real time Green's functions. In a companion paper we elucidate a generic scheme for perturbative calculations of interacting models, with particular reference to the Anderson model.

\end{abstract}
\maketitle

\section{Introduction}
Quantum systems exhibiting an interplay of interactions and out-of-equilibrium effects are of significant interest and constitute an active area of research.\cite{Goldhaber,Kontos,Egger,Yacoby,LeHur_Frac,Mora_LeHur,Bockrath_2010,McEuen,Mason,Mirlin} In contrast to the situation of equilibrium physics, however, there is currently no unifying theoretical framework for describing the dynamics of a generic quantum system out of equilibrium. The real-time Schwinger-Keldysh formalism\cite{Schwinger,Keldysh,Rammer} has been successful in the treatment of specific systems,\cite{Meir_Wingreen}  but occasionally gives rise to pathological perturbative expansions which suffer from infrared divergences.\cite{Parcollet_Hooley} This can be viewed as an artifact of the infinite contour used in the integration, and one generally requires additional relaxation mechanisms to regulate the theory. While quantum impurity models in equilibrium have been extensively studied,\cite{Hewson} probing and understanding properties of nonequilibrium steady states is a far more subtle task, many aspects of which are yet to be explored. Nevertheless, there have been significant advances in our understanding via several distinct approaches, which include the scattering Bethe Ansatz,\cite{Mehta_Andrei,Mehta_Andrei_new} field theory techniques,\cite{Fendley,Boulat_2008,doyon_new_2007} time-dependent density matrix renormalization group (RG),\cite{Boulat_2008,Dagotto} time-dependent numerical RG,\cite{Anders_2008,Anders_2010} perturbative RG,\cite{Rosch,Chung,schoeller1} Hamiltonian flow equations,\cite{Kehrein} functional RG,\cite{Schoeller,Karrasch,Wolfle} strong-coupling expansions,\cite{Mora_2009,LeHur_2008,Mora_2008,Mitra} diagrammatic Monte Carlo,\cite{Fabrizio,Millis,Eckstein,Muehlbacher} and imaginary-time nonequilibrium quantum Monte Carlo.\cite{Han_2007,Han_2010}

Mesoscopic quantum objects such as quantum dots (``artificial atoms''), characterized by a set of discrete energy levels, in contact with reservoir leads,\cite{Glazman} are described by quantum impurity models. In this context, the Anderson model, which describes a single discrete level coupled via tunneling to a Fermi-liquid sea, is of particular interest.\cite{Hewson} For a level with sufficiently low energy and with strong on-site interaction, the Anderson model mimics the situation of a Coulomb-blockaded quantum dot.\cite{Devoret} In this regime, charge degrees of freedom on the level are frozen out and the Anderson model becomes intimately related to the Kondo model. The latter describes a magnetic impurity (manifested by the spin of the highest occupied level for an odd number of electrons on the quantum dot) entangled with the spins of a Fermi sea. (In the mesoscopic setting, the Fermi sea is replaced by the reservoir leads.) The Kondo entanglement then produces a prominent Abrikosov-Suhl resonance in the density of states of the quantum dot at the Fermi level,
which results in perfect transparency when the quantum dot is symmetrically coupled to its leads
(the source and the drain).\cite{Goldhaber_1998,Delft,Weis,Glazman,schoeller2} Nanoscale systems can be routinely driven out of equilibrium by applying a bias voltage between the two reservoir leads. Among the various issues which arise out of equilibrium, the precise fate of the Abrikosov-Suhl resonance\cite{Leturcq} when applying a finite bias voltage, remains a delicate issue. In general, it seems essential to elaborate theoretical and numerical methods which will allow access to the full current-voltage characteristics of interacting mesoscopic systems.

The present paper constitutes Part I of a two-part series. Its main purpose is to reformulate electronic nonequilibrium transport in quantum impurity models in terms of an effective equilibrium steady-state density matrix. This concept was initiated by Hershfield,\cite{hershfield_reformulation_1993,Hershfield_Schiller} who proposed that the appropriate density matrix in Boltzmann form could be constructed explicitly by invoking the Lippmann-Schwinger operators of the system. Recently, there have been several numerical efforts at implementing this scheme \cite{Anders_2008,Anders_2010,Han_2007,Han_2010}. However, the theoretical foundations of this approach were yet to be firmly established. In this work, we concretely establish the validity of this effective equilibrium approach by making use of the open system limit, \cite{Doyon_Andrei}which guarantees that the system relaxes to a steady state. This well-controlled mathematical contrivance mimics the role of relaxation mechanisms, and does not require the explicit inclusion of the bath degrees of freedom at the level of the Hamiltonian. Furthermore, we demonstrate the equivalence between observables computed using the Schwinger-Keldysh framework and the effective equilibrium approach. It is important to emphasize the fact that the effective equilibrium description encompasses the case when one includes interactions on the dot. We then propose an imaginary-time formulation within this framework, which allows us to build general formulae and establish the mathematical machinery for evaluating quantities relevant to transport, and to systematically implement perturbative and non-perturbative techniques familiar from finite temperature equilibrium field theory. 
In Part II,\cite{Dutt_2} we employ the formalism in the context of the nonequilibrium Anderson model and perform a systematic perturbative expansion in the repulsive interaction on the dot. Other aspects of the method, including numerical approaches, have been recently investigated in Refs. \onlinecite{Han_2006,Han_2007,Han_2010,Oguri_2007,doyon_new_2007, Anders_2008,Anders_2010}. In addition, we note that the non-interacting resonant level model may also be a pertinent starting point in tackling situations with strong interactions, as for example the large-$N$ Anderson model\cite{Read_Newns,Mitra,Han_2006} or the Toulouse limit of the Kondo model.\cite{Toulouse}

Our paper is organized as follows. In Section II, we introduce the model and define the ``open-system limit''.\cite{Doyon_Andrei,Mehta_Andrei_new} This concept provides a transparent framework that guarantees the existence and uniqueness of a steady-state density matrix and we will use it recurrently in our discussion. In Section III, we present the effective equilibrium approach and justify the validity of the method. We provide an alternative derivation of the effective equilibrium density matrix proposed by Hershfield\cite{hershfield_reformulation_1993} and establish its equivalence with the formal expression for the density matrix proposed by Doyon and Andrei \cite{Doyon_Andrei}.
\ Furthermore, we present useful recursion relations for the expectation values of
observables which underpin the equivalence between this approach and the time dependent prescription. In Section IV, we demonstrate the equivalence between observables computed within this framework and the Schwinger-Keldysh formalism which enables us to arrive at the Meir-Wingreen formula for the current\cite{Meir_Wingreen_1,Meir_Wingreen_2}. We then derive a compact expression for the charge occupation on the dot, whereby we identify the spectral function of the dot as the central quantity of interest. It is instructive to note that this effective equilibrium formalism avoids the (often complex) coupled Dyson's equations for the various Green's functions on the Keldysh contour, and it is in principle possible to compute the Green's function of interest directly. In Section V 
we develop the methodology appropriate for the effective equilibrium approach in terms of a finite temperature imaginary-time formalism, which can be used to analytically compute real time Green's functions.  

\section{Model}
\begin{figure}
    \centering
        \includegraphics[width=0.8\columnwidth]{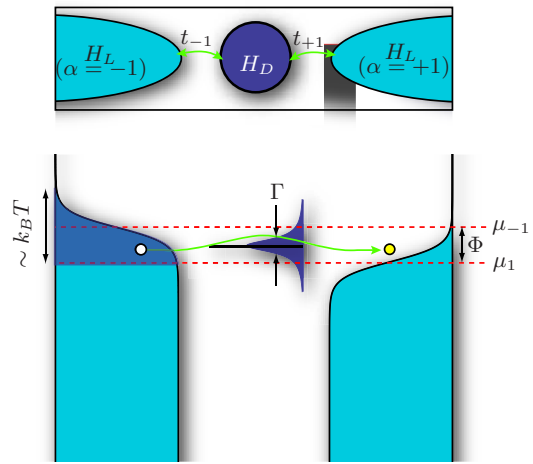}
        \caption{Schematic setup (upper panel) and energy diagram (lower panel) of a generic quantum impurity model out of equilibrium. The system is at a temperature $T$, and $\Phi=\mu_{1}-\mu_{-1}\equiv eV$ denotes the voltage bias between the source and drain leads. The (bare) energy of the dot level is given by $\epsilon_{d}$. The energy broadening of this level, given by the width $\Gamma$, is due to the tunnel coupling between dot and leads.\label{fig:transport}}    
\end{figure}

As a prototype for our discussion of Hershfield's approach we consider a system consisting of two Fermi-liquid leads, coupled by tunnel junctions to a central system with a number of discrete levels, the ``dot", see Fig.\ \ref{fig:transport}. The Hamiltonian of such a generic system can be written in the form $H=H_{L}+H_{D}+H_{T}$. The first term
\be
H_{L}=\sum_{\alpha k \sigma}\epsilon_{\alpha k}c^{\dag}_{\alpha k\sigma}c_{\alpha k\sigma}
\ee
describes the left and right leads ($\alpha=\pm1$), where $c_{\alpha k\sigma}$ ($c_{\alpha k\sigma}^\dag)$ annihilates (creates) an electron (strictly speaking a Fermi-liquid quasiparticle) in state $k$ with spin projection $\sigma$ in lead $\alpha$. The corresponding energy dispersion is denoted by $\epsilon_{\alpha k}$. These leads couple to the dot, whose Hamiltonian is given by
\be
H_{D}=\ \sum_{\sigma}\epsilon_{d}d^{\dag}_{\sigma}d_{\sigma}+H_\text{int}
\ee
with $d_{\sigma}$ ($d_{\sigma}^\dag$) annihilating (creating) an electron with spin $\sigma$ in the  discrete level with energy $\epsilon_d$. Any electron interactions are lumped into the contribution $H_\text{int}$, and we will assume that these interactions are localized on the dot (as it is the case for the Anderson model). Finally, the tunneling of electrons between leads and dot is captured by the tunneling Hamiltonian
\be
H_{T}=\frac{1}{\sqrt{\Omega}}\sum_{\alpha k \sigma}t_{\alpha k}\left(c^{\dag}_{\alpha k\sigma}d_{\sigma}+\text{h.c.}\right),
\label{eq:tunneling}
\ee
where $\Omega$ is the lead volume (assumed identical for both leads) and $t_{\alpha k}$ specifies the tunneling matrix element for electron transfer between state $k$ in lead $\alpha$ and the discrete dot state.
In the presence of a bias voltage, realized as a chemical potential difference  $\Phi=\mu_{1}-\mu_{-1}$ between left and right lead, the tunneling induces an electric current. 

Typically, the steady state is reached after a short time determined by relevant relaxation rates in the leads. For the purpose of calculations aiming at steady-state quantities, it is convenient to avoid the consideration of microscopic relaxation mechanisms and instead invoke the so-called ``open-system limit". \cite{Doyon_Andrei,Mehta_Andrei_new} In short, this approach proceeds as follows:
Initially, up to some time $t=t_0<0$ in the early past, the tunneling term is absent and the Hamiltonian of the system is given by $H_{0}=H_{L}+H_{D}$, such that leads and dot are decoupled. For $t<t_0$ the system is hence described by the separable density matrix
\begin{align}
	\rho_{0}=\exp\left[-\beta\left(H_{0}-\frac{\Phi}{2}\sum_{\alpha}\alpha N_{\alpha}\right)\right].
\end{align}
Here, $\beta=(k_B T)^{-1}$ denotes the inverse temperature and  $N_{\alpha}=\sum_{k\sigma}c^{\dag}_{\alpha k\sigma}c_{\alpha k\sigma}$ is the number operator of electrons in lead $\alpha$. Between times $t_0<t<0$, the tunneling is then `switched on' adiabatically, {\it i.e.}, \ $H=H_{0}+H_{T}e^{\eta t}\theta(t-t_{0})$, where the parameter $\eta\rightarrow 0^{+}$ defines a slow switch-on rate.  At time $t=0$ the tunneling has reached its full strength, the system is in its steady state and observables can be evaluated.

As demonstrated in Refs.\ \onlinecite{Doyon_Andrei} and \onlinecite{Mehta_Andrei_new}, the existence and uniqueness of steady state is tied to the validity of the inequalities $v_F/L \ll |t_0|^{-1} \ll \eta$, where $v_F$ denotes the Fermi velocity and $L$ the linear system size. Intuitively, these inequalities ensure that hot electrons hopping onto a given lead at time $t_0$ will not be reflected back and return to the junction before the measurement process, and further that the process of switching on $H_T$ remains adiabatic. We also note that the energy scale of switch-on $|t_0|^{-1}$ suffices to smear out the energy level spacing $v_F/L$. In this sense, the openness of the system provides the ``dissipation'' mechanism necessary for the steady state, allowing the high-energy electrons to escape to infinity and thus, effectively relax.
On a more technical level, the inequalities result in the factorization of long-time correlation functions, which facilitates the proof of existence and uniqueness of steady state. 
 
The crucial ingredient in our discussion is that $L/v_F$ determines the largest time scale in our problem. The exact protocol by which we switch on the tunneling is irrelevant for the formation of steady state. Let us take as an example the non-interacting case, and for the sake of convenience assume that we have identical leads, {\it i.e.}, $t_{\alpha k}=t$. Here, instead of adiabatically turning on the tunneling suppose we do a quench, one observes that the transients decay with a relaxation time $\sim\Gamma$, where $\Gamma=2\pi t^{2}\nu$ denotes the linewidth of the dot.\cite{Thomas_2008} Here $\nu$ symbolizes the density of states, which for simplicity is assumed to be a  constant. To make the argument concrete, assume that the tunneling Hamiltonian is absent for $t<0$, and the distribution functions in the leads are given by Fermi functions $f(\ep-\al\frac{\Phi}{2})$ ($\alpha=+1,-1$ specify the source and drain reservoirs respectively). Furthermore, suppose that the dot is initially unoccupied, {\it i.e.} $n_{d}^{\text{in}}=0$ denotes the initial charge on the dot. The occupation of the dot at a time t($>0$) is given by\cite{Thomas_2008}
 \begin{align}
n_{d}(t)&=n_{d_{\text{ss}}}\left(1+e^{-2 \Gamma t}\right)-\frac{2\Gamma e^{-\Gamma t}}{\pi}\notag\\
&\times\int d\omega\left[f(\ep-\frac{\Phi}{2})+f(\ep+\frac{\Phi}{2})\right]\frac{\text{cos}\left[(\omega-\ep_{d})t\right]}{\Gamma^{2}+(\omega-\ep_{d})^{2}}.
\end{align}
Here
\begin{align}
n_{d_{\text{ss}}}=\frac{\Gamma}{\pi}\int d\omega\frac{f(\ep-\frac{\Phi}{2})+f(\ep+\frac{\Phi}{2})}{\Gamma^{2}+(\omega-\ep_{d})^{2}}
\label{eq:ssoccupation}
\end{align}
denotes the steady state expectation value of the dot occupation, after the transients have decayed. It is interesting to note that if we turn the coupling of the QD to the left and right leads adiabatically to zero at the same rate, (formally this implies taking $\Gamma\rightarrow 0$ in Eq. (\ref{eq:ssoccupation})) then the final QD occupation is given by
\begin{align}
n_{d}^{\text{fin}}=f(\ep_{d}-\frac{\Phi}{2})+f(\ep_{d}+\frac{\Phi}{2}).
\end{align} 
The fact that $n_{d}^{\text{in}}\neq n_{d}^{\text{fin}}$, clearly illustrates the irreversibility of the turning on process. It encapsulates precisely how the system is driven out of equilibrium and is the reason we are required to analytically continue time to the Keldysh contour. This demonstrates that the steady state formalism cannot be continued from zero to finite tunneling perturbatively. Therefore, in our effective equilibrium approach, we include the contribution of the tunneling terms non-perturbatively. Interactions on the other hand can be turned on adiabatically. This procedure is reversible and does not exhibit similar anomalies. 

\section{Derivation of the Effective Equilibrium Form of the Steady-State Density Matrix}
The description of a nonequilibrium problem in terms of an effective equilibrium density matrix was first proposed by Hershfield in Ref.\ \onlinecite{hershfield_reformulation_1993}. To facilitate such a description, the nonequilibrium steady-state density matrix $\rho$ is rewritten in the usual Boltzmann form,
\begin{equation}\label{rhoHershfield}
\rho=\exp\left[-\beta(H - Y)\right],
\end{equation}
at the cost of introducing a correction operator, which we will call (following Hershfield's convention) the $Y$ operator.\footnote{Note that throughout the text, density matrices are not assumed to be normalized. Normalization is always introduced explicitly when evaluating expectation values.} Formally, the definition of this correction operator as 
\be
Y=\frac{1}{\beta}\ln \rho+ H
\ee
 is always possible. However, in this form it is neither particularly elucidating nor useful for calculating nonequilibrium transport properties. Hershfield put forward the idea that $Y$ can be expressed explicitly and compactly in terms of Lippmann-Schwinger operators\cite{gell-mann_formal_1953,Lippmann}  $\psi_{\alpha k\sigma}$,
which are fermionic operators that diagonalize the full Hamiltonian,\cite{Han_2006}
\begin{align}
H=\sum_{\alpha k}\epsilon_{\alpha k}\psi^{\dag}_{\alpha k\sigma}\psi_{\alpha k\sigma}.
\end{align}
Hershfield proposed that the $Y$ operator has the general form\cite{hershfield_reformulation_1993}
\begin{equation}
Y=\frac{\Phi}{2}\sum_{\alpha k\sigma}\alpha\psi^{\dag}_{\alpha k\sigma}\psi_{\alpha k\sigma}.\label{Yrep}
\end{equation}
Since this operator encodes the entire $\Phi$ dependence, the $Y$ operator is also called bias operator. Note that $Y$ vanishes at zero bias and the steady-state density matrix correctly simplifies to the equilibrium density matrix. 

In this Section we provide a detailed proof of the representation of $Y$ given in Eq.\ \eqref{Yrep}. In contrast to Hershfield, we do not invoke the presence of additional relaxation mechanisms to reach steady state. Instead, we make systematic use of the time-dependent open system approach,\cite{Doyon_Andrei,Mehta_Andrei_new} which circumvents ill-defined expressions and makes the proof rigorous.

\subsection{Proof of the explicit form of the $Y$ operator}
The structure of the proof is as follows. The starting point is the expansion of the steady-state density matrix into a power series in the tunneling Hamiltonian $H_T$. This can be accomplished either by using the Hershfield form of the density matrix, or by employing the time-dependent framework of the open-system limit where the steady-state density matrix is obtained by adiabatically switching on the coupling in the far past. In both cases, one obtains analytical expressions for the power series in $H_T$. Order for order comparison of these expansions then leads to a system of nested differential equations for the bias operators. Finally, the solution to this system of differential equations is then shown to be identical with the representation of $Y$ in terms of Lippmann-Schwinger operators, see Eq.\ \eqref{Yrep}. We should emphasize the fact that the expansion in powers of $H_{T}$ is a purely formal procedure, which we adopt for the sake of a systematic comparison, and the proof below is non-perturbative in $H_{T}$.

\subsubsection{Expansion of $\rho$ using the effective equilibrium representation}
In the derivation of the explicit form of the $Y$ operator, it is convenient to collect terms in orders of the tunneling Hamiltonian $H_{T}$. We thus start by expanding $Y=\sum_{n=0}^{\infty}Y_{n}$ into a series in powers of the tunneling, such that $Y_{n}\propto \left(H_{T}\right)^{n}$. In the following, the index $n$ will always be used for power counting of the tunneling Hamiltonian $H_{T}$. From Eq.\ \eqref{rhoHershfield} we thus obtain 
\begin{align}
\rho&=\exp\left[ -\beta\left\{(H_{0} - Y_{0})+(H_{T}-Y_{1})-\sum_{n=2}^{\infty}Y_{n}\right\}\right]\notag\\
&=\exp\left[-\beta \sum_{n=0}^{\infty}X_{n}\right].
\end{align}
Here, we have regrouped the Hamiltonian with the $Y$ operator order by order in the auxiliary operator $X_n\sim (H_T)^n$, which  is hence defined as
\be\label{XY}
X_n \equiv \begin{cases}
H_0 - Y_0, & n=0\\
H_T - Y_1, & n=1\\
- Y_n,     & n\ge2\ .
\end{cases}
\ee
 
One can expand the exponential operator to collect terms in powers of the tunneling $H_{T}$ such that 
\begin{align}
\rho&
=\sum_{l=0}^{\infty}\frac{(-\beta)^{l}}{l!}\sum_{i_{1},\ldots,i_{l}=0}^{\infty} X_{i_{1}}\ldots X_{i_{l}}\notag\\
&=\sum_{n=0}^{\infty}\sum_{l=0}^{\infty}\frac{(-\beta)^{l}}{l!}\sum_{i_{1}+\cdots+i_{l}=n} X_{i_{1}}\ldots X_{i_{l}} \equiv\sum_{n=0}^{\infty}\rho_{n},
\label{rho-exp}
\end{align}
where, following our general notation, $\rho_n$ denotes the order $(H_T)^n$ contribution to the steady-state density matrix.

\subsubsection{Expansion of $\rho$ using the open-system approach}
Let us now employ the open-system approach (as outlined in Section II). In this case, the steady-state density matrix  is obtained by switching on the tunneling in the early past $t_0<0$. Up to this time $t_0$, the leads are decoupled from the dot. The adiabatic switch-on of tunneling is facilitated by using $H_{T}e^{\eta t}\theta(t-t_{0})$ for the tunneling Hamiltonian (note that in this scenario, the total Hamiltonian is therefore time dependent). At time $t=0$, transients have decayed and time evolution has turned the original density matrix $\rho_0=\bar{\rho}(t=t_0)$ into the steady-state density matrix $\rho=\bar{\rho}(t=0)$.  We note that the condition of adiabaticity is strictly true only in the limit $\frac{1}{\vert t_{0}\vert}\ll\eta$, where $\vert t_{0}\vert\rightarrow \infty$, which is assumed in the open-system limit.\cite{Mehta_Andrei_new} We emphasize that the actual evaluation of this limit is deferred until the very end of all calculations. Keeping $1/|t_{0}|$ and $\eta$ small but nonzero in the interim is crucial for mathematical clarity and for avoiding ill-defined expressions.

This time, it will be convenient to work in the interaction picture (with respect to the tunneling), where
in general 
\be
{\cal O}_{I}(t)=e^{iH_{0}(t-t_{0})}{\cal O}e^{-iH_{0}(t-t_{0})}.
\ee
denotes the interaction picture of the operator $\mathcal{O}$. Note that the time $t_0$ (not $t=0$) has been chosen as the reference time where Schr\"odinger and interaction pictures agree, $\mathcal{O}=\mathcal{O}_I(t_0)$. [In this we differ from the conventions adopted by Hershfield,\cite{hershfield_reformulation_1993} who chose different reference times for the Heisenberg and interaction representation.] 

In the interaction picture, the density matrix satisfies the evolution equation
\begin{equation}
i\frac{d}{dt}\bar{\rho}_{I}(t)=[H_{T,I}(t),\bar{\rho}_{I}(t)].
\label{eq:densitymatrix}
\end{equation}
This is formally solved by
\begin{align}\label{eq:interactionsolution}
\bar{\rho}_{I}(t)&=\sum_{n=0}^{\infty}\frac{(-i)^n}{n!}{\cal T}\bigg\{\prod_{i=1}^{n}\left[\int_{t_{0}}^{t}dt_{i}\right]\\\notag
&\qquad\times[H_{T,I}(t_{1}),[H_{T,I}(t_{2}),[\ldots[H_{T,I}(t_{n}),\rho_{0}]]\ldots]\bigg\}\\\nonumber
&\equiv \sum_{n=0}^\infty \bar{\rho}_{n,I}(t),
\end{align}
where the $n=0$ term simply fixes the boundary condition $\bar{\rho}(t=t_0)=\rho_0$ and $\mathcal{T}$ denotes time-ordering of operators.

It should be noted that Eq.\ \eqref{eq:interactionsolution} already has the form of a power series in the tunneling and thus has been used to define the $n$-th order contribution $\bar{\rho}_{n,I}(t)$ to the interaction-picture density matrix. For later purposes it is useful to note that the contributions can alternatively be obtained from
\begin{align}
\frac{d}{dt}\bar{\rho}_{n,I}(t)=-i[H_{T,I}(t),\bar{\rho}_{n-1,I}(t)],
\label{eq:nested}
\end{align}
i.e., a system of nested differential equations associated with the boundary conditions $\bar{\rho}_{0,I}(t_0)=\rho_0$ and $\bar{\rho}_{n,I}(t_0)=0$ for $n\ge1$.

\subsubsection{Derivation of differential equations for $Y_n$}
For the results to be consistent, we require that the steady-state density matrix $\rho$ be identical to the steady-state density matrix obtained in the open-system limit. This allows one to determine the correct form of the $Y$ operator, order for order in the tunneling. 

We thus impose the identity of the steady-state density matrices
\begin{equation}
e^{iH_{0}(t-t_{0})}\rho_{n}e^{-iH_{0}(t-t_{0})}=\bar{\rho}_{n,I}(t), 
\label{eq:interactionpic}
\end{equation}
where we have transformed $\rho_n$ into the interaction picture. Eq.\ \eqref{eq:interactionpic} is expected to hold for all $t>0$, since at that point the interaction has been fully switched on, and the time-independent Hamiltonian (used in Hershfield's effective equilibrium approach) and the time-dependent Hamiltonian (from the open-system approach) are identical.

We now differentiate the left-hand side of Eq.\ \eqref{eq:interactionpic} with respect to time and use Eq.\ \eqref{rho-exp} to obtain
\begin{align}
\label{eq:diff1}
&\frac{d}{dt}\rho_{n,I}(t)=\\\nonumber
&\sum_{l=1}^{\infty}\frac{(-\beta)^{l}}{l!}\sum_{i_{1}+\cdots+i_{l}=n}\; 
\sum_{k=1}^{l} X_{i_{1},I}\cdots\frac{dX_{i_{k},I}(t)}{dt}\cdots X_{i_{l},I},
\end{align}
where $X_{i_{k},I}=e^{iH_{0}(t-t_{0})}X_{i_{k}}e^{-iH_{0}(t-t_{0})}$ denotes the interaction picture representation of $X_{i_{k}}$.
Similarly, we may differentiate the right-hand side of Eq.\ \eqref{eq:interactionpic}. The resulting commutator is given in Eq.\eqref{eq:nested} and contains $\rho_{n-1,I}(t)$,\footnote{Note that due to the identity \eqref{eq:interactionpic} we may, from here on, drop all bars on $\rho$.} for which we substitute the corresponding expression from Eq.\ \eqref{rho-exp}. This way, we obtain
\begin{widetext}
\begin{align}
\frac{d}{dt} \rho_{n,I}(t)=i[\rho_{n-1,I}(t),H_{T,I}(t)]
&=i\sum_{l=1}^{\infty}\frac{(-\beta)^{l}}{l!}\sum_{i_{1}+i_{2}+\ldots+i_{l}=n-1}
\;\sum_{k=1}^{l} X_{i_{1},I}(t)\ldots[X_{i_{k},I}(t),H_{T,I}(t)]\ldots X_{i_{l},I}(t)\notag\\\label{eq:diff2}
&=i\sum_{l=1}^{\infty}\frac{(-\beta)^{l}}{l!}\sum_{i_{1}+i_{2}+\ldots+i_{l}=n}
\;\sum_{k=1}^{l}X_{i_{1},I}(t)\ldots[X_{i_{k}-1,I}(t),H_{T,I}(t)]\ldots X_{i_{l},I}(t),
\end{align}
\end{widetext}
where for $n<0$ we define $X_n=0$.
In the last step of Eq.\ \eqref{eq:diff2}, the summation constraint is shifted from $n-1$ to $n$ to facilitate the comparison with Eq.\ \eqref{eq:diff1}. This comparison yields the relation
\begin{align}
\frac{d}{dt}X_{n,I}=i[X_{n-1,I}(t),H_{T,I}(t)].
\end{align}
Finally, utilizing the relation \eqref{XY} between $X$ and $Y$ operators, one finds that the $Y_{n}$ operators also satisfy
\begin{align}
\frac{d}{dt}Y_{n,I}=i[Y_{n-1,I}(t),H_{T,I}(t)].
\label{eq:Yoperatordiff}
\end{align}
To obtain the $Y$ operator from these differential equations, it is crucial to specify the boundary conditions at the initial time $t=t_0$. Given the relation $e^{iH_{0}(t-t_{0})}e^{\cal O} e^{-iH_{0}(t-t_{0})}=e^{{\cal O}_{I}(t)}$, valid for any operator ${\cal O}$, the boundary condition $\bar{\rho}(t=t_{0})=\rho_{0}$ implies \cite{Doyon_Andrei}
\be\label{boundcond}
\lim_{t\searrow t_{0}}Y_{I}(t)=\frac{\Phi}{2}\sum_{\alpha}\alpha N_{\alpha}.
\ee
It now remains to prove that the following interaction-picture expression of $Y$
 \begin{align}
 Y_{I}(t)=&\frac{\Phi}{2}\sum_{\alpha k \sigma}\alpha \psi^{\dag}_{\alpha k\sigma,I}(t)\psi_{\alpha k\sigma,I}(t),
\label{eq:Y operator}
\end{align}
in terms of the Lippmann-Schwinger operators $\psi^{\dag}_{\alpha k\sigma,I}(t)$ represents the solution to the above initial-value problem. Once we have recapitulated the crucial properties of these Lippmann-Schwinger operators in the following subsection, it will be simple to confirm that this ansatz indeed solves the differential equation \eqref{eq:Yoperatordiff} subject to the boundary condition \eqref{boundcond}.

\subsubsection{Properties of the Lippmann-Schwinger operators}
It has been shown~\cite{Han_2006} for the generic model defined in Section II that the Lippmann-Schwinger operators are fermionic,
\begin{align}
\left\{\psi^\dagger_{\alpha k\sigma},\psi_{\alpha'k'\sigma'}\right\}=\delta_{\alpha\alpha'}\delta_{kk'}\delta_{\sigma\sigma'},
\label{eq:LSoperator}
\end{align}
and diagonalize the full Hamiltonian (including all interactions),
\begin{align}
H=\sum_{\alpha k}\epsilon_{\alpha k}\psi^{\dag}_{\alpha k\sigma}\psi_{\alpha k\sigma}.
\end{align}
One can expand $\psi^{\dag}_{\alpha k\sigma}$ in powers of the tunneling Hamiltonian $H_{T}$, {\it i.e.}, 
\be
\psi^{\dag}_{\alpha k\sigma}=\sum_{n=0}^{\infty}\psi^{\dag}_{\alpha k\sigma,n}
\ee
such that $\psi^{\dag}_{\alpha k\sigma,n}\propto \left(H_{T}\right)^{n}$. In the interaction representation, the operators $\psi^{\dag}_{\alpha k\sigma}$ satisfy
\begin{align}
\frac{d}{dt}\psi^{\dag}_{\alpha k\sigma,I}(t)&=i[H_{0},\psi^{\dag}_{\alpha k\sigma,I}(t)]\notag\\
=i[H_{I}(t),&\psi^{\dag}_{\alpha k\sigma,I}(t)]-i[H_{T,I}(t),\psi^{\dag}_{\alpha k\sigma,I}(t)],
\label{eq:diffeqnth}
\end{align}
subject to the boundary condition 
\be
\lim_{t\searrow t_0} \psi_{\alpha k \sigma,I}(t) = c_{\alpha k \sigma}.
\ee
Equation \eqref{eq:diffeqnth} can be further simplified, and collecting orders of $(H_{T})^{n}$, cast into the set of nested differential equations
\be
\frac{d}{dt}\psi^{\dag}_{\alpha k\sigma,n,I}(t)=i\epsilon_{\alpha k}\psi^{\dag}_{\alpha k\sigma,n,I}(t)+i[\psi^{\dag}_{\alpha k\sigma,n-1,I}(t),H_{T,I}(t)]. 
\ee

As discussed in Ref.\ \onlinecite{Han_2006}, the formal solution can be expressed compactly as
\begin{equation}
\psi^\dagger_{\alpha k\sigma}=c^\dagger_{\alpha k\sigma}
+\frac{1}{\epsilon_{\alpha k}-{\cal L}+i\eta}{\cal L}_{T} 
c^\dagger_{\alpha k\sigma}
\label{eq:lippmannformal}
\end{equation}
in terms of Liouvillian superoperators ${\cal L}$ and ${\cal L}_{T}$. The action of such superoperators on any operator $\mathcal{O}$ is defined as  ${\cal L}_{A}{\cal O}=[A,\mathcal{O}]$. Further, $\eta\rightarrow0^{+}$ is a regularization similar in spirit to the one utilized in the more familiar Lippmann-Schwinger \emph{states} in scattering theory. From Eq.\ \eqref{eq:lippmannformal} one can write the detailed form of the operator $\psi^{\dag}_{\alpha k\sigma,n}$
\begin{align}
\psi^\dagger_{\alpha k\sigma}=\sum_{n=0}^{\infty}\left[\left(\frac{1}{\epsilon_{\alpha k}-{\cal L}_{0}+i\eta}{\cal L}_{T}\right)^{n}c^\dagger_{\alpha k\sigma}\right]\equiv\sum_{n}\psi^\dagger_{\alpha k\sigma,n}.
\end{align}
Note that when the tunneling is set to zero, one finds indeed that $\psi^\dagger_{\alpha k\sigma}\rightarrow c^\dagger_{\alpha k\sigma}$.

\subsubsection{Conclusion of the proof}
It is now simple to verify that the ansatz for $Y_{n,I}(t)$ solves the set of differential equations given by Eq.\ (\ref{eq:Y operator}), and obeys the appropriate boundary condition \eqref{boundcond}: We start by writing the $(H_T)^n$ contribution to the $Y$ operator as
\begin{align}
Y_{n,I}(t)=\sum_{\alpha k\sigma}\alpha\frac{\Phi}{2}\sum_{p=0}^{n}\psi^{\dag}_{\alpha k\sigma,p,I}(t)\psi_{\alpha k\sigma,n-p,I}(t).
\end{align}
Differentiating this with respect to time yields
\begin{align}
&\frac{d}{dt}Y_{n,I}(t)=\sum_{\alpha k\sigma}\alpha\frac{\Phi}{2}\sum_{p=0}^{n}\bigg[\left(\frac{d}{dt}\psi^{\dag}_{\alpha k\sigma,p,I}(t)\right)\psi_{\alpha k\sigma,n-p,I}(t)\notag\\
&\qquad\qquad\qquad+\psi^{\dag}_{\alpha k\sigma,p,I}(t)\left(\frac{d}{dt}\psi_{\alpha k\sigma,n-p,I}(t)\right)\bigg],
\end{align}
so that in conjunction with Eq.\ (\ref{eq:diffeqnth}) one obtains
\begin{align}
&\frac{d}{dt}Y_{n,I}(t)=\sum_{\alpha k\sigma}\alpha\frac{\Phi}{2}\sum_{p=0}^{n-1}[\psi^{\dag}_{\alpha k\sigma,p,I}(t)\psi_{\alpha k\sigma,n-1-p,I}(t)]\notag\\
&\qquad\qquad\qquad=i[Y_{n-1,I}(t),H_{t,I}].
\end{align}
This concludes the proof.

In summary, we find that the steady state dynamics of the system can be described by an effective equilibrium density matrix of the form
\begin{align}
{\rho}&=e^{-\beta(H - Y)},
\end{align}
where the operator
\begin{align}
Y=\frac{\Phi}{2}\sum_{\alpha k \sigma}\alpha\psi^{\dag}_{\alpha k\sigma}\psi_{\alpha k\sigma}
\end{align}
encodes the entire nonequilibrium boundary condition of the system.

\subsection{Recursion relations for expectation values of observables}
In addition to the previous proof, we follow Hershfield\cite{hershfield_reformulation_1993} and underpin the equivalence of the adiabatic approach and the effective equilibrium approach by showing that they lead to identical recursion relations for expectation values of observables. Again, the systematic use of the open-system limit makes the proof sound.

Let ${\cal O}$ denote a generic observable. To derive the first recursion relation, we will decompose the steady-state expectation value $\langle{\cal O}\rangle$ into a series counting the powers of the tunneling Hamiltonian $H_T$. In the first step, we thus substitute the expansion \eqref{rho-exp} of $\rho$,
\begin{align}
\langle{\cal O}\rangle&=\frac{\Tr[\rho{\cal O}]}{\Tr[\rho]}
=\frac{\Tr\left[\sum_{n=0}^{\infty}\rho_{n}{\cal O}\right]}{\Tr\left[\sum_{m=0}^{\infty}\rho_{m}\right]}.
\end{align}
Pulling out a factor of $1/\Tr[\rho_0]$, expanding the denominator as a geometric series, and collecting terms order for order in $H_T$, we can rewrite this as
\begin{align}
\langle{\cal O}\rangle=&\sum_{n=0}^{\infty} \sum_{l=0}^{n}(-1)^{l}\sideset{}{'}\sum_{j_{1},\ldots,j_{l}=1}^\infty \prod_{s=1}^l
\left[\frac{\Tr[\rho_{j_{s}}]}{\Tr[\rho_{0}]}\right]
\frac{\Tr[\rho_{n-\sum_{s=1}^l j_{s}}{\cal O}]}{\Tr[\rho_{0}]}\nonumber\\
\equiv&\sum_{n=0}^{\infty}\langle\mathcal{O}\rangle_{n},\label{on-eq}
\end{align}
where $\sum'$ denotes a restricted summation, subject to the condition $\sum_{s=1}^l j_{s}\le n$. Eq. \eqref{on-eq} allows one to prove the important recursion relation
\begin{align}
\langle{\cal O}\rangle_{n}=\frac{\Tr[\rho_{n}{\cal O}]}{\Tr[\rho_{0}]}-\sum_{k=1}^{n}\frac{\Tr[\rho_{k}]}{\Tr[\rho_{0}]}\langle{\cal O}\rangle_{n-k},
\label{eq:timeindep}
\end{align}
which relates the $n$-th order term to the expectation value of $\mathcal{O}$ with respect to $\rho_n$, and all lower-order terms $\langle{\cal O}\rangle_{m}$ ($m=0,\ldots,n-1$).

Now, we turn to the time-dependent representation assuming an adiabatic switch-on of the tunneling.
The steady-state expectation value of an observable may now be obtained via
\begin{equation}
\langle {\cal O}\rangle=\frac{\Tr[\bar{\rho}_{I}(0){\cal O}_{I}(0)]}{\Tr[\rho_{0}]},\label{oi-exp}
\end{equation}
where we have switched to the interaction picture.
To arrive at the desired recursion relation, we again decompose this into a power series of the tunneling Hamiltonian,
\be
\langle {\cal O}\rangle=\frac{\Tr[\bar{\rho}_{I}(0){\cal O}_{I}(0)]}{\Tr[\rho_{0}]} = \sum_{n=0}^\infty \langle {\cal O}\rangle_n,
\ee 
where
\begin{align}\label{t-on}
&\langle{\cal O}\rangle_{n}=\frac{(-i)^n}{n!}\frac{1}{\Tr\rho_{0}}\Tr\bigg\{{\cal T}\prod_{i=1}^{n}\left[\int_{t_{0}}^{t=0}dt_{i}\right]\\\notag
&\quad\times[H_{T,I}(t_{1}),[H_{T,I}(t_{2}),[\ldots[H_{T,I}(t_{n}),\rho_{0}]]\ldots]{\cal O}_{I}(0)\bigg\}.
\end{align}	
This expression for $\langle{\cal O}\rangle_{n}$ allows one to derive the second recursion relation, which reads
\begin{align}
&\langle {\cal O}\rangle_{n}=\frac{\Tr[\rho_{n,I}(0){\cal O}_{I}(0)]}{\Tr[\rho_{0}]} -\sum_{k=1}^{n}\frac{\Tr[\rho_{k,I}(0)]}{\Tr[\rho_{0}]}\langle{\cal O}\rangle_{n-k}.
\label{eq:timedep}
\end{align}
The details of this derivation are given in Appendix \ref{app:1}. The proof makes explicit use of the factorization of two-time correlation functions in the limit of large time separation.\cite{Doyon_Andrei} The agreement between Eqs. (\ref{eq:timeindep}) and (\ref{eq:timedep}) underpins the equivalence between the time-dependent adiabatic approach and the effective equilibrium approach.

\section{Correspondence with the Schwinger-Keldysh Approach\label{sec:correspondence}}
Having obtained the effective equilibrium form of the steady state density matrix, we now illustrate how one can compute transport observables such as the current and the spectral function within this description. We propose an imaginary time formulation for treating such nonequilibrium systems in a manner similar to finite temperature field theory.

\noindent 
In imaginary time, we define the propagation of an operator by
\begin{align}
{\cal O}(\tau)&=e^{\tau(H-Y)}{\cal O}e^{-\tau(H-Y)}=e^{\tau({\cal L}-{\cal L}_{Y})}{\cal O}.
\end{align}
The nonequilibrium thermal Green's function is defined on $0<\tau<\beta$ as
\begin{align}
{\cal G}_{{\cal O}_{1}{\cal O}_{2}}(\tau)=-\langle{\cal T}\left[{\cal O}_{1}(\tau){\cal O}_{2}(0)\right]\rangle=-\langle{\cal O}_{1}(\tau){\cal O}_{2}(0)\rangle.
\end{align}
Fourier transforming in imaginary time this results in
\begin{align}
{\cal G}_{{\cal O}_{1}{\cal O}_{2}}(i\omega_{n})=\left\langle\left\{{\cal O}_{1},\frac{e^{i\omega_{n}0^{{+}}}}{i\omega_{n}-{\cal L}+{\cal L}_{Y}}{\cal O}_{2}\right\}\right\rangle,
\end{align}
where $\omega_{n}=(2n+1)\pi/\beta$ ($n\in {\mathbb Z}$) denotes the fermionic Matsubara frequencies.

Switching to real time, the Heisenberg representation of an operator ${\cal O}$ is given by ${\cal O}(t)=e^{iHt}{\cal O}e^{-iHt}$. The nonequilibrium real time retarded Green's function can then be expressed as 
\begin{align}
G^{\text{ret}}_{{\cal O}_{1}{\cal O}_{2}}(t)&=-i\theta(t)\langle\{{\cal O}_{1}(t),{\cal O}_{2}(0)\}\rangle\notag\\
&=-i\theta(t)\frac{\Tr\left[e^{-\beta(H-Y)}\left\{{\cal O}_{1}(t),{\cal O}_{2}(0)\right\}\right]}{\Tr\left[e^{-\beta(H-Y)}\right]}.
\end{align}
By using the spectral representation and then Fourier transforming, we obtain from this
\be
G^{\text{ret}}_{{\cal O}_{1}{\cal O}_{2}}(\omega)=\left\langle\left\{{\cal O}_{1},\frac{1}{\omega-{\cal L}+i\eta}{\cal O}_{2}\right\}\right\rangle.
\ee
We emphasize that, despite the effective equilibrium character of the Hershfield approach, there does remain one important difference between the effective equilibrium description and any regular equilibrium many-body theory. This distinction arises from the different propagators  in imaginary versus real time; namely, in Hershfield's effective equilibrium formalism all Heisenberg operators in imaginary time evolve under the modified Hamiltonian $H-Y$, whereas Heisenberg operators in real time  evolve under the Hamiltonian $H$. As a result, imaginary time and real time Green's function, ${\cal G}_{{\cal O}_{1}{\cal O}_{2}}(i\omega_{n})$ and $G^{\text{ret}}_{{\cal O}_{1}{\cal O}_{2}}(\omega)$  are not simply related by a direct analytic continuation $i\omega_{n}\rightarrow\omega+i\eta$.

As an observable of prime interest in transport, let us now consider the current flowing through the dot. It is given by
\begin{align}
I&=\frac{I_{1}+I_{-1}}{2}=-\frac{e}{2}\left\langle\frac{d\left(N_{1}(t)-N_{-1}(t)\right)}{dt}\right\rangle\notag\\
&=-\frac{e}{2}\sum_{\alpha}\alpha\left\langle\frac{dN_{\alpha}(t)}{dt}\right\rangle=i\frac{e}{2}\sum_{\alpha}\alpha\left\langle[N_{\alpha}(t),H]\right\rangle\notag\\
&=i\sum_{\alpha k\sigma}\alpha \frac{et_{\alpha k}}{2\sqrt{\Omega}}\left\langle\left(c^{\dag}_{\alpha k\sigma}d_{\sigma}-d^{\dag}_{\sigma}c_{\alpha k\sigma}\right)\right\rangle\notag\\
&=i\sum_{\alpha k\sigma}\alpha \frac{et_{\alpha k}}{2\sqrt{\Omega}}\left({\cal G}_{d_{\sigma}c^{\dag}_{\alpha k\sigma}}(\tau=0)-{\cal G}_{c_{\alpha k\sigma}d^{\dag}_{\sigma}}(\tau=0)\right)\notag\\
&=\iim\left[\sum_{\alpha k \sigma}\alpha\frac{et_{\alpha k}}{\sqrt{\Omega}}{\cal G}_{c_{\alpha k \sigma}d_{\sigma}^\dag}(\tau=0)\right].
\label{current_primary}
\end{align}
where $-e$ denotes the charge of the electron.
Here, we have used the fact that the system is in a steady state and the current is unchanged under time translations. As we shall see below, it will prove to be most convenient to express the current in terms of the Fourier representation of the imaginary time Green's function, namely
\begin{align}
 I&=\iim\left[\sum_{\alpha k \sigma\omega_{n}}\alpha \frac{e t_{\alpha k}}{\sqrt{\Omega}}\frac{1}{\beta}{\cal G}_{c_{\alpha k \sigma}d_{\sigma}^\dag}(i\omega_{n})\right].
\label{eq:current1}
\end{align}

We now show the equivalence of this approach with the Schwinger-Keldysh formalism and recover the familiar Meir-Wingreen formula for the steady-state current.\cite{Meir_Wingreen_1,Meir_Wingreen_2} 
For simplicity, let us assume that the coupling to the dot is independent of $k$, \emph{i.e.} $t_{\alpha k}=t_{\alpha}$, and that the leads are identical $\epsilon_{\alpha k}=\epsilon_{k}$. (This is not a strict requirement and the proof can be easily generalized.) Our starting point is Eq.\ \eqref{eq:current1} and we now write explicitly
\begin{align}
&\frac{1}{\beta}\sum_{\omega_{n}}{\cal G}_{c_{\alpha k \sigma}d_{\sigma}^\dag}(i\omega_{n})=\frac{1}{\beta}\sum_{\omega_{n}}\left\langle\left\{c_{\alpha k \sigma},\frac{e^{i\omega_{n}0^{{+}}}}{i\omega_{n}-{\cal L}+{\cal L}_{Y}}d^{\dag}_{\sigma}\right\}\right\rangle.
\end{align}
Using Eq.\ \eqref{eq:lippmannformal} for $c_{\alpha k\sigma}$ and evaluating the effect of $\mathcal{L}_T$, we obtain
\begin{align}
&\frac{1}{\beta}\sum_{\omega_{n}}{\cal G}_{c_{\alpha k \sigma}d_{\sigma}^\dag}(i\omega_{n})=\frac{1}{\beta}\sum_{\omega_{n}}\bigg[\left\langle\left\{\psi_{\alpha k \sigma},\frac{e^{i\omega_{n}0^{{+}}}}{i\omega_{n}-{\cal L}+{\cal L}_{Y}}d^{\dag}_{\sigma}\right\}\right\rangle\notag\\
&-\left\langle\left\{\frac{1}{\epsilon_{k}+{\cal L}-i\eta}d_{\sigma},\frac{e^{i\omega_{n}0^{{+}}}}{i\omega_{n}-{\cal L}+{\cal L}_{Y}}d^{\dag}_{\sigma}\right\}\right\rangle\bigg].\label{aux1}
\end{align}
With a small trick, one can show that the second term does not need to be evaluated when computing the current $I$. To see this, we exploit the fact that in steady state, the current in the left and right junctions must be identical and hence, we can write the current as a weighted average of the form  $I=\left(t_{1}^{2}I_{-1}+t_{-1}^{2}I_{1}\right)/\left(t_{1}^{2}+t_{-1}^{2}\right)$. This eliminates the presence of the second term in Eq.\ \eqref{aux1} from the expression for the current. Proceeding with the remaining term we obtain in several steps,
\begin{widetext}
\begin{align}
&\frac{1}{\beta}\sum_{\omega_{n}}\left\langle\left\{\psi_{\alpha k \sigma},\frac{e^{i\omega_{n}0^{{+}}}}{i\omega_{n}-{\cal L}+{\cal L}_{Y}}d^{\dag}_{\sigma}\right\}\right\rangle=\frac{1}{\beta}\sum_{\omega_{n}}\left\langle\left\{\frac{e^{i\omega_{n}0^{{+}}}}{i\omega_{n}+{\cal L}-{\cal L}_{Y}}\psi_{\alpha k \sigma},d^{\dag}_{\sigma}\right\}\right\rangle\notag\\
&\qquad\qquad\qquad\qquad=\frac{1}{\beta}\sum_{\omega_{n}}\left\langle\left\{\frac{e^{i\omega_{n}0^{{+}}}}{i\omega_{n}-\epsilon_{k}+\alpha\Phi/2}\psi_{\alpha k \sigma},d^{\dag}_{\sigma}\right\}\right\rangle=f\left(\epsilon_{k}-\alpha\frac{\Phi}{2}\right)\langle\{\psi_{\alpha k\sigma},d^{
\dag}_{\sigma}\}\rangle\notag\notag\\
&\qquad\qquad\qquad\qquad=\frac{t_{\alpha}}{\sqrt{\Omega}}f\left(\epsilon_{k}-\alpha\frac{\Phi}{2}\right)\left\langle\left\{\frac{1}{\epsilon_{k}+{\cal L}-i\eta}d_{\sigma},d^{\dag}_{\sigma}\right\}\right\rangle=\frac{t_{\alpha}}{\sqrt{\Omega}}f\left(\epsilon_{k}-\alpha\frac{\Phi}{2}\right) G^\text{ret}_{d_{\sigma}d_{\sigma}^\dag}(\epsilon_{k}),
\label{part1}
\end{align}
\end{widetext}
where the transition from the first to the second line is facilitated by the general relation \eqref{appb-eq}. Also, in going from the sum over $k$ to the integral over $\epsilon_{k}$ we make the replacement $\frac{1}{\Omega}\sum_{k}\rightarrow \int \nu\,d\epsilon_{k}$. Here we have followed the standard procedure and assumed the spectrum has been linearized. Further $\nu$ denotes the density of states, which is assumed to be a constant. 
Finally, this results in the well-known Meir-Wingreen expression for the current,\cite{Meir_Wingreen_1,Meir_Wingreen_2}
\begin{align}
I=&2e\frac{\Gamma_{1}\Gamma_{-1}}{\Gamma_{1}+\Gamma_{-1}}\int d\epsilon_{k}A_{d}(\epsilon_{k})\notag\\
&\qquad\qquad\times\left[f\left({\epsilon_{k}+\frac{\Phi}{2}}\right)-f\left({\epsilon_{k}-\frac{\Phi}{2}}\right)\right],
\label{eq:meirwingreen}
\end{align}
where 
\begin{align}
A_{d}(\epsilon_{k})&=-\frac{1}{\pi}\sum_{\sigma}\iim \left[G^{\text{ret}}_{d_{\sigma}d_{\sigma}^\dag}(\epsilon_{k})\right]
\end{align}
 is the nonequilibrium spectral function of the dot and is general a function of the bias voltage $\Phi$. We shall however suppress this explicit bias dependence of the spectral function. In the above expression $\Gamma_{\alpha}=\pi t_{\alpha}^{2}\nu$ denotes the partial broadening of the level due to the coupling to the lead $\alpha$. 

Let us now derive the above identities via a slightly different route and recover the traditional form of the Meir Wingreen formula. Proceeding as in Eq.\ \eqref{aux1} one observes
\begin{align}
&\frac{1}{\beta}\sum_{\omega_{n}}{\cal G}_{d_{\sigma} c_{\alpha k \sigma}^\dag}(i\omega_{n})=\frac{1}{\beta}\sum_{\omega_{n}}\bigg[\left\langle\left\{d_{\sigma},\frac{e^{i\omega_{n}0^{{+}}}}{i\omega_{n}-{\cal L}+{\cal L}_{Y}}\psi^{\dag}_{\alpha k \sigma}\right\}\right\rangle\notag\\
&-\left\langle\left\{\frac{e^{i\omega_{n}0^{{+}}}}{i\omega_{n}+{\cal L}-{\cal L}_{Y}}d_{\sigma},\frac{1}{\epsilon_{k}-{\cal L}+i\eta}d^{\dag}_{\sigma}\right\}\right\rangle\bigg].\label{aux2}
\end{align}
It is simple to show, in a sense similar to Eq.\ \eqref{part1} that the 1st part of the above expression
\begin{align}
&\frac{1}{\beta}\sum_{\omega_{n}}\left\langle\left\{d_{\sigma},\frac{e^{i\omega_{n}0^{{+}}}}{i\omega_{n}-{\cal L}+{\cal L}_{Y}}\psi^{\dag}_{\alpha k \sigma}\right\}\right\rangle\notag\\
&\qquad\qquad\qquad=\frac{t_{\alpha}}{\sqrt{\Omega}}f\left(\epsilon_{k}-\alpha\frac{\Phi}{2}\right)G^{\text{adv}}_{d_{\sigma}d^{\dag}_{\sigma}}(\epsilon_{k}).
\label{part2}
\end{align}
Thus, using Eq.\ \eqref{current_primary} together with Eqs.\ \eqref{part1} and \eqref{part2} we get 
\begin{widetext}
\begin{align}
I&=i\sum_{\alpha k\sigma}\alpha \frac{et_{\alpha k}}{2\sqrt{\Omega}}\left({\cal G}_{d_{\sigma}c^{\dag}_{\alpha k\sigma}}(\tau=0)-{\cal G}_{c_{\alpha k\sigma}d^{\dag}_{\sigma}}(\tau=0)\right)\notag\\
&=i\frac{e}{2\pi}\sum_{\alpha\sigma}\alpha\Gamma_{\alpha}\int_{-\infty}^{\infty} d\epsilon_{k}\bigg\{f\left(\epsilon_{k}-\alpha\frac{\Phi}{2}\right)\left[G^{\text{adv}}_{d_{\sigma}d^{\dag}_{\sigma}}(\epsilon_{k})-G^{\text{ret}}_{d_{\sigma}d^{\dag}_{\sigma}}(\epsilon_{k})\right]\notag\\
&\qquad\qquad-\frac{1}{\beta}\sum_{\omega_{n}}\bigg[\left\langle\left\{\frac{e^{i\omega_{n}0^{{+}}}}{i\omega_{n}+{\cal L}-{\cal L}_{Y}}d_{\sigma},\frac{1}{\epsilon_{k}-{\cal L}+i\eta}d^{\dag}_{\sigma}\right\}\right\rangle-\left\langle\left\{\frac{1}{\epsilon_{k}+{\cal L}-i\eta}d_{\sigma},\frac{e^{i\omega_{n}0^{{+}}}}{i\omega_{n}-{\cal L}+{\cal L}_{Y}}d^{\dag}_{\sigma}\right\}\right\rangle\bigg]\bigg\}.
\end{align}
Using an identity from Appendix \ref{app:2}, while integrating over $\epsilon_{k}$ in the last 2 terms, we find that
\begin{align}
I&=i\frac{e}{2\pi}\sum_{\alpha\sigma}\alpha\Gamma_{\alpha}\bigg\{\int_{-\infty}^{\infty} d\epsilon_{k} f\left(\epsilon_{k}-\alpha\frac{\Phi}{2}\right)\left[G^{\text{adv}}_{d_{\sigma}d^{\dag}_{\sigma}}(\epsilon_{k})-G^{\text{ret}}_{d_{\sigma}d^{\dag}_{\sigma}}(\epsilon_{k})\right]+2i\pi {\cal G}_{d_{\sigma}d^{\dag}_{\sigma}}(\tau=0)\bigg\}.
\end{align}
Now $2i\pi {\cal G}_{d_{\sigma}d^{\dag}_{\sigma}}(\tau=0)=2i\pi\left\langle d^{\dag}_{\sigma}d_{\sigma}\right\rangle=2\pi G^{<}(t=0)=\int_{-\infty}^{\infty}d\epsilon_{k} G^{<}(\epsilon_{k})$. This helps us recover the traditional form of the Meir Wingreen formula:
\begin{align}
I&=i\frac{e}{2\pi}\sum_{\alpha\sigma}\alpha\Gamma_{\alpha}\int_{-\infty}^{\infty} d\epsilon_{k} \bigg\{f\left(\epsilon_{k}-\alpha\frac{\Phi}{2}\right)\left[G^{\text{adv}}_{d_{\sigma}d^{\dag}_{\sigma}}(\epsilon_{k})-G^{\text{ret}}_{d_{\sigma}d^{\dag}_{\sigma}}(\epsilon_{k})\right]+ G^{<}(\epsilon_{k})\bigg\}.
\end{align}
\end{widetext}
 
We proceed in a similar fashion to derive an expression for the charge occupation on the dot
\begin{align}
n_{d}&=\sum_{\sigma}\langle d^{\dag}_{\sigma}d_{\sigma}\rangle=\frac{1}{\beta}\sum_{\sigma,\omega_{n}}{\cal G}_{d_{\sigma}d_{\sigma}^\dag}(i\omega_{n})\notag\\
&=\frac{1}{2}\int d\epsilon_{k} A_{d}(\epsilon_{k})\left[f\left(\epsilon_{k}-\frac{\Phi}{2}\right)+f\left(\epsilon_{k}+\frac{\Phi}{2}\right)\right]\notag\\
&=\int d\epsilon_{k} A_{d}(\epsilon_{k})f^{\text{eff}}(\epsilon_{k},\Phi),
\label{eq:occupation}
\end{align}
where 
\begin{align}
f^{\text{eff}}(\epsilon_{k},\Phi)=\frac{1}{2}\left[f\left(\epsilon_{k}-\frac{\Phi}{2}\right)+f\left(\epsilon_{k}+\frac{\Phi}{2}\right)\right].
\end{align}
Note that on taking the limit $\Phi\rightarrow 0$, Eq.\eqref{eq:occupation} reduces to a form well-known from equilibrium many-body theory. In Appendix\ \ref{app:2} we provide a straightforward derivation of this result using the spectral representation within the effective equilibrium formulation. This result can also be obtained within the Keldysh framework in a slightly indirect manner as we have shown in Appendix\ \ref{app:3}.  
This perhaps accounts for its omission in the existing literature, where the charge occupation is typically obtained using the lesser Green's function $G^{<}$.

We thus identify the non-equilibrium spectral function $
(\omega,\Phi)$ as the central quantity of interest, which one can use to calculate transport observables. In the effective equilibrium scheme this can be done directly, without resorting to the Dyson equations which couple the various Green's functions, and can be complicated. It is instructive to note that for the non-interacting case the spectral function is independent of the bias, but the introduction of interactions generally imparts to the spectral function a complex bias dependence.

\section{Generating functional framework}
As usual, the evolution in imaginary time of an operator in Heisenberg picture is governed by the propagator $e^{-\tau(H-Y)}$, which closely matches the form of the density matrix $e^{-\beta(H-Y)}$. This similarity makes the imaginary time description very convenient for the evaluation of Green's functions when using functional techniques. We will now outline the general methodology for constructing Green's functions using the effective equilibrium density matrix within a functional framework. 

We construct the  generating functional as a coherent state functional integral over Grassmann variables, corresponding to the Lippmann-Schwinger operators
\begin{align}
Z[J,J^{\ast}]=\int\prod_{\alpha k \sigma}\left({\cal D} \psi^{\ast}_{\alpha k \sigma } {\cal D} \psi_{\alpha k \sigma}\right)e^{-S[J,J^\ast]},
\end{align}
where $S$ denotes the action
\begin{align}
S&=\int_{0}^{\beta}d\tau\sum_{\alpha k \sigma }\bigg[\psi^{\ast}_{\alpha  k \sigma }(\partial_{\tau}+\epsilon_{\alpha k\sigma}-\alpha\Phi/2)\psi_{\alpha k \sigma }\notag\\
&\qquad\qquad\qquad\quad-\big(J^{\ast}_{\alpha k \sigma }\psi_{\alpha k \sigma }+\psi^{\ast}_{\alpha k \sigma }J_{\alpha k \sigma }\big)\bigg],
\end{align}
and for notational convenience, we have suppressed the fact that the Grassmann variables are functions of imaginary time.
Any Green's function in terms of the Lippmann-Schwinger operators is now given by an appropriate functional derivative of the generating functional
\begin{align}
{\cal G}_{\psi_{\alpha'k'\sigma'}\psi_{\alpha k\sigma}^\dag}(\tau)&=-\langle{\cal T}[\psi_{\alpha k\sigma}(\tau)\psi^{\dag}_{\alpha'k'\sigma'}(0)]\rangle\notag\\
=\frac{1}{Z_{0}}&\frac{\delta^{2}}{\delta J^{\ast}_{\alpha k\sigma}(\tau)\delta J_{\alpha'k'\sigma'}(0)}Z(J,J^{\ast})\bigg\vert_{J=0}.
\label{eq:functionald}
\end{align}
Here $Z_{0}=Z(0)$ denotes the generating functional with the source terms absent.

We observe that the generating functional and consequently the Green's functions have a trivial form in terms of the Lippmann-Schwinger operators. However for transport we have to compute Green's functions in terms of the original degrees of freedom of the system such as ${\cal G}_{d_{\sigma}d_{\sigma}^\dag}$ and ${\cal G}_{c_{\alpha k\sigma}d_{\sigma}^\dag}$. The goal now is to express the $c_{\alpha k\sigma}$ and the $d_{\sigma}$ in terms of these Lippmann-Schwinger operators. This is by no means a trivial task, since the precise form of the Lippmann-Schwinger operators is in general not known. However for a system governed by single particle dynamics, the exact form of the scattering states is readily calculated and the actual degrees of freedom of the system and these states bear a linear relationship to each other. In this limit, it is straightforward to pass from one set of operators to the other, and it will constitute the starting point of our calculations. Later this will lay the foundation for computing Green's functions in interacting theories, perturbatively in the interaction.

For systems governed by single particle dynamics the Lippmann-Schwinger operators are given as a linear combination of lead electron and dot electron operators 
\begin{align}
\psi^{\dag}_{\alpha k\sigma}=\sum_{\alpha'k'\sigma'}\Lambda_{\alpha'k'\sigma'}c^{\dag}_{\alpha'k'\sigma'}+\sum_{\sigma'}\kappa_{\sigma'}d^{\dag}_{\sigma'},
\end{align}
where $\Lambda_{\alpha'k'\sigma'}$ and $\kappa_{\sigma'}$ are appropriate c-numbers. 

For interacting theories, the form of the Lippmann-Schwinger operator turns out to be quite non-trivial and can be written schematically as 
\begin{align}
&\psi^{\dag}_{\varsigma}=\sideset{}{'}\sum_{\upsilon_{i},\xi_{j},\upsilon'_{k},\xi'_{l}}\Lambda_{\upsilon_{1}\ldots\upsilon'_{1}\ldots\xi_{1}\ldots\xi'_{1}\ldots}\notag\\
&\times\prod_{i}\left(c_{i}^{\dag}\right)^{\upsilon_{i}}\prod_{j}\left(d_{j}^{\dag}\right)^{\xi_{j}}\prod_{k}\left(c_{k}\right)^{\upsilon'_{k}}\prod_{l}\left(d_{l}\right)^{\xi'_{l}},
\end{align}
where the summation $\sum'$ is over $\upsilon_{i},\xi_{j},\upsilon'_{k},\xi'_{l}\in\{0,1\}$ with the constraint $\sum_{i}\upsilon_{i}+\sum_{j}\xi_{j}-\sum_{k}\upsilon'_{k}-\sum_{l}\xi'_{l}=1$, so that in total there is one more creation operator than there are annihilation operators. Here the coefficients $\Lambda_{\upsilon_{1}\ldots\upsilon'_{1}\ldots\xi_{1}\ldots\xi'_{1}\ldots}$ are c-numbers and we have used an abbreviated notation $i,j,k,l$ for the labels of the single-particle lead and dot states. Thus, in a generic interacting model a simple way to pass between the Lippmann-Schwinger basis and the original basis is in general not easily accomplished. The construction of the Lippmann-Schwinger operator order by order in the interaction is however well defined and from this we can compute the Green's functions in powers of the interaction strength, as shown in the companion paper (part II). For the purposes of numerics, it is possible to circumvent explicit construction of the Lippmann-Schwinger operator by working at the level of the density matrix.\cite{Anders_2010}

The functional formalism described above provides a convenient platform for the implementation of non-perturbative techniques. We are currently using this scheme for a large-$N$ treatment of the infinite-$U$ Anderson model out of equilibrium, following the slave-boson language of Read and Newns \cite{Read_Newns, Mitra}. The saddle point solution of the functional integral in this case, mimics the non-interacting theory and it is possible to systematically build corrections in powers of $1/N$. 

\section{Conclusions}
In this paper, we have provided a rigorous alternative derivation of the effective equilibrium density matrix approach proposed by Hershfield, in the context of nonequilibrium quantum transport using the concept of an ``open system''. Furthermore, we have illustrated the methodology of this formulation for a quantum dot, in which the reservoir leads play the role of good (infinite) thermal baths, such that a unique steady state exists. The latter can be cast into the form of an effective equilibrium density matrix, where the associated modified Hamiltonian can be explicitly written in terms of the Lippmann-Schwinger scattering state operators. 

We have detailed the foundations of the theory and demonstrated a rigorous correspondence between the observables computed using the Schwinger-Keldysh formalism and the effective equilibrium approach
. We emphasize that an advantage of the effective
equilibrium approach is that it is conducive to the implementation of numerical methods, such as numerical RG\cite{Anders_2008,Anders_2010}
and Quantum Monte Carlo.\cite{Han_2010}
Furthermore, we have built a generating functional framework using an imaginary-time formulation that will constitute the basic mathematical machinery for our computation of the steady-state dynamics of the system.
In particular,\cite{Dutt_2} we will develop an infrared-convergent systematic perturbation theory scheme to treat interactions within the effective equilibrium approach. This scheme enables a direct computation of the non-equilibrium spectral function, from which we can infer transport observables, without resorting to (often complicated) coupled Dyson's equations which arise in the Keldysh context.

Finally, we underline that the non-interacting resonant level model 
may serve as an appropriate starting point in the case with strong interactions, {\it e.g.}, in the large-$N$ Anderson model \cite{Read_Newns, Mitra} or in the close vicinity of the Toulouse limit in the Kondo model.\cite{Toulouse} The effective equilibrium method may also be generalized to more sophisticated systems, such as one-dimensional mesoscopic systems embodied by the Luttinger paradigm\cite{Yacoby,LeHur_Frac} where it is possible, in principle, to combine bosonization and the scattering picture.\cite{Hershfield_Schiller}

\begin{acknowledgments}
 We acknowledge N.\ Andrei and H.\ Baranger for stimulating discussions. This work is supported by the
Department of Energy under grant DE-FG02-08ER46541 (P.D. and K.L.H.), by the Yale Center for Quantum Information Physics (NSF DMR-0653377, J.K. and K.L.H.) and by the NSF under grant DMR-0907150 (J.E.H.).
\end{acknowledgments}

\appendix
\section{Derivation of the recursive form of ${\cal O}_{n}$}
\label{app:1}
In this appendix we provide the detailed derivation of the recursion relation \eqref{eq:timedep} when working within the time-dependent approach, switching on the tunneling adiabatically. Our starting point for the derivation is the expression for the order-$(H_T^n)$ contribution to the steady-state expectation value of some operator $\mathcal{O}$,
\begin{align}
\langle {\cal O}\rangle_{n}=&\frac{(-i)^{n}}{\Tr[\rho_{0}]}\Tr\bigg\{\int_{t_{0}}^{0}dt_{n}\int_{t_{0}}^{t_{n}}dt_{n-1}\cdots\int_{t_{0}}^{t_{2}}dt_{1}\notag\\
[H_{T,I}(t_{n})&,[H_{T,I}(t_{n-1}),[\ldots[H_{T,I}(t_{1}),\rho_{0}]]\ldots]{\cal O}_{I}(t=0)\bigg\},
\label{eq:nthterm}
\end{align}
see Eq.\ \eqref{t-on} in the main text. Carrying out the $t_{1}$ integration and using Eq.\ (\ref{eq:densitymatrix}), one obtains
\begin{align}
&\langle {\cal O}\rangle_{n}=\frac{(-i)^{n-1}}{\Tr[\rho_{0}]}\Tr\bigg[\int_{t_{0}}^{0}dt_{n}\int_{t_{0}}^{t_{n}}dt_{n-1}\cdots\int_{t_{0}}^{t_{3}}dt_{2}\\
&\times[H_{T,I}(t_{n}),[\ldots[H_{T,I}(t_{2}),\left(\rho_{1,I}(t_{2})-\rho_{1,I}(t_{0})\right)]\ldots]{\cal O}_{I}(0)\bigg].\notag
\end{align}
We now retain the term involving $\rho_{1,I}(t_{0})$, and proceed by carrying out the $t_{2}$ integration for the term containing $\rho_{1,I}(t_{2})$. This integration yields 2 terms: one with $\rho_{2,I}(t_{0})$ which we retain, and the other with $\rho_{2,I}(t_{3})$ which we subject to further integration. Continuing in this fashion until all the variables $t_{1},\ldots,t_{n}$ in the latter term have been integrated out, we arrive at the result 
\begin{align}
&\langle{\cal O}\rangle_{n}=\frac{\Tr\left[\rho_{n,I}(0){\cal O}_{I}(0)\right]}{\Tr[\rho_{0}]}\\\notag
&-\sum_{p=1}^{n}\frac{(-i)^{n-p}}{\Tr[\rho_{0}]}\int_{t_{0}}^{t_{n+1}=0}dt_{n}\int_{t_{0}}^{t_{n}}dt_{n-1}\cdots\int_{t_{0}}^{t_{p+2}}dt_{p+1}\notag\\
&\times\Tr\bigg\{[H_{T,I}(t_{n}),[\ldots,[H_{T,I}(t_{p+1}),\rho_{p,I}(t_{0})]\ldots]{\cal O}_{I}(0)\bigg\}.
\end{align}
Let us focus on the $p$-th term of the above sum. One observes that when $p=n$, none of the integrals are present and the argument of the trace is just $\rho_{p,I}(t_{0}){\cal O}_{I}(0)$. Suppressing the integrals and other c-numbers we isolate the product of operators within the trace 
\begin{align}
&\Tr\bigg\{[H_{T,I}(t_{n}),[\ldots,[H_{T,I}(t_{p+1}),\rho_{p,I}(t_{0})]]\ldots]{\cal O}_{I}(0)\bigg\}\notag.\\
\intertext{We can cycle $\rho_{p,I}(t_{0})$ in the above expression out of the nested commutators, thereby inverting the nested commutator structure to re-express this as}
&\Tr\bigg\{\rho_{p,I}(t_{0})[\ldots[{\cal O}_{I}(0),H_{T,I}(t_{n})],...H_{T,I}(t_{p+1})]\bigg\}.
\end{align}
This allows us to rewrite $\langle\mathcal{O}\rangle_n$ in the form
\begin{align}\label{eq:form1}
&\langle{\cal O}\rangle_{n}=\frac{\Tr\left[\rho_{n,I}(0){\cal O}_{I}(0)\right]}{\Tr[\rho_{0}]}\\\notag
&-\sum_{p=1}^{n}\frac{(-i)^{n-p}}{\Tr[\rho_{0}]}\int_{t_{0}}^{0}dt_{n}\int_{t_{0}}^{t_{n}}dt_{n-1}\cdots\int_{t_{0}}^{t_{p+2}}dt_{p+1}\\\notag
&\times\Tr\bigg[\rho_{p,I}(t_{0})[\ldots[{\cal O}_{I}(0),H_{T,I}(t_{n})],\ldots,H_{T,I}(t_{p+1})]\bigg].
\end{align}
To proceed further, let
\begin{align}
\langle\langle\cdots\rangle\rangle_{0}=\frac{\Tr\left(e^{-\beta(H_{0}-Y_{0})}\cdots\right)}{\Tr\left(e^{-\beta(H_{0}-Y_{0})}\right)},
\end{align}
denote an expectation value with respect to the density matrix $\rho_{0}$. It has been shown by Doyon and Andrei in Ref.\ \onlinecite{Doyon_Andrei} that, in the open system limit, long-time correlation functions factorize, \emph{i.e.}
\begin{equation}
\langle\langle{\cal O}(t_{1}){\cal O}(t_{2})\rangle\rangle_{0}\rightarrow\langle\langle{\cal O}(t_{1})\rangle\rangle_{0}\langle\langle{\cal O}(t_{2})\rangle\rangle_{0}
\label{eq:factorize}
\end{equation}
as $\vert t_{1}-t_{2}\vert\rightarrow\infty$. We will now employ this factorization in Eq.\ (\ref{eq:form1}). Renaming the integration variables $t'_{1}\equiv t_{p+1}$, $t'_{2}\equiv t_{p+2}$, $\ldots$, $t'_{n-p}\equiv t_{n}$ and inserting $\rho_{0}\rho_{0}^{-1}=\openone$ we get
\begin{widetext}
\begin{align}
&\langle{\cal O}\rangle_{n}=\frac{\Tr\left[\rho_{n,I}(0){\cal O}_{I}(0)\right]}{\Tr[\rho_{0}]}\\\notag
&-\sum_{p=1}^{n}\frac{(-i)^{n-p}}{\Tr[\rho_{0}]}\int_{t_{0}}^{0}dt'_{n-p}\int_{t_{0}}^{t'_{n-p}}dt'_{n-p-1}\cdots\int_{t_{0}}^{t'_{2}}dt'_{1}
\Tr\bigg\{\rho_{0}\underbrace{\rho_{0}^{-1}\rho_{p,I}(t_{0})}_{A(t_{0})}\underbrace{[[\ldots[{\cal O}_{I}(0),H_{T,I}(t'_{n-p})],\ldots],H_{T,I}(t'_{1})]}_{B(t'_{1},\ldots,t'_{n-p},t=0)}\bigg\}.
\label{eq:factor1}
\end{align}

We now recall that in the open system limit we take $v_{F}/L\ll1/\vert t_{0}\vert\ll\eta$ with $t_{0}\rightarrow-\infty$. This implies that the $e^{\eta t}$ term in $H_{T,I}$ essentially cuts off the lower limit of the time integrals at a time $\sim1/\eta$ much earlier than $t_{0}$. Thus, Eq.\ \eqref{eq:factor1} can be cast into an integral over $\langle\langle A(t_{0})B(t'_{1},\ldots,t'_{n-p},t=0)\rangle\rangle_{0}$, where $\vert t'_{1}-t_{0}\vert$,$\ldots$,$\vert t'_{n-p}-t_{0}\vert$,$\vert 0-t_{0}\vert\rightarrow\infty$. Using Eq.\ (\ref{eq:factorize}) we get
\begin{align}
&\langle\langle A(t_{0})B(t'_{1},\ldots,t'_{n-p},t=0)\rangle\rangle_{0}=\langle\langle A(t_{0})\rangle\rangle_{0}\langle\langle B(t'_{1},\ldots,t'_{n-p},t=0)\rangle\rangle_{0},
\end{align}
and hence
\begin{align}
&\langle{\cal O}\rangle_{n}=\frac{\Tr\left[\rho_{n,I}(0){\cal O}_{I}(0)\right]}{\Tr[\rho_{0}]}\\\notag
&-\sum_{p=1}^{n}\frac{\Tr[\rho_{p,I}(t_{0})]}{\Tr[\rho_{0}]}\Tr\bigg[\big(-i)^{n-p}\int_{t_{0}}^{0}dt'_{n-p}\int_{t_{0}}^{t'_{n-p}}dt'_{n-p-1}
\cdots\int_{t_{0}}^{t'_{2}}dt'_{1}[[\ldots[{\cal O}_{I}(0),H_{T,I}(t'_{n-p})]\ldots,]H_{T,I}(t'_{1})]{\rho}_{0}\bigg].\notag
\end{align}
Inverting the nested commutator structure as previously, we get
\begin{align}
\langle{\cal O}\rangle_{n}&=\frac{\Tr\left[\rho_{n,I}(0){\cal O}_{I}(0)\right]}{\Tr[\rho_{0}]}\\\notag
&-\sum_{p=1}^{n}\frac{\Tr[\rho_{p,I}(t_{0})]}{\Tr[\rho_{0}]}\Tr\bigg[\big(-i)^{n-p}\int_{t_{0}}^{0}dt'_{n-p}\int_{t_{0}}^{t'_{n-p}}dt'_{n-p-1}
\cdots\int_{t_{0}}^{t'_{2}}dt'_{1}[H_{T,I}(t'_{n-p}),[\ldots,[H_{T,I}(t'_{1}),{\rho}_{0}]]\ldots]{\cal O}_{I}(0)\bigg].
\end{align}
\end{widetext}
Finally, comparing the second term in the latter equation with Eq.\ (\ref{eq:nthterm}) and noting that $\Tr[\rho_{p,I}(t_{0})]=\Tr[\rho_{p,I}(0)]$, we arrive at the final recursion relation
\begin{align}
\langle{\cal O}\rangle_{n}=\frac{\Tr\left[\rho_{n,I}(0){\cal O}_{I}(0)\right]}{\Tr[\rho_{0}]}-\sum_{p=1}^{n}\frac{\Tr[\rho_{p,I}(0)]}{\Tr[\rho_{0}]}\langle{\cal O}\rangle_{n-p}.
\end{align}

\section{Derivation of $n_{d}$ using the spectral representation}
\label{app:2}
In this appendix we provide additional details on the equivalence between the effective equilibrium approach and the Schwinger-Keldysh formalism and explicitly show the identity of the electron occupancy on the quantum dot as obtained in the Schwinger-Keldysh formalism and as obtained with the effective equilibrium approach. We start by stating two useful identities involving Liouvillian superoperators.
The first identity regards the imaginary-time Heisenberg representation of an operator ${\cal O}$ and is given by
\begin{align}
 {\cal O(\tau)}=e^{\tau(H-Y)}{\cal O}e^{-\tau(H-Y)}=e^{\tau({\cal L}-{\cal L}_{Y})}\mathcal{O}.
\end{align}
The second identity is a small trick which transfers a Liouvillian superoperator from one side of an anticommutator to the other,
\begin{align}
\left\langle \left\{  {\cal O}_{1}, \frac{1}{z-{\cal L}}{\cal O}_{2} \right\} \right\rangle
=\left\langle  \left\{  \frac{1}{z+{\cal L}}{\cal O}_{1}, {\cal O}_{2} \right\} \right\rangle,
\label{appb-eq}
\end{align}
where $z$ is a c-number.

We now set out to demonstrate the equivalence of the Schwinger-Keldysh expression for the dot occupancy,
\begin{align}\label{nd-keldysh}
n_{d}&=\frac{1}{2}\sum_{\sigma}\int_{-\infty}^{\infty}d\epsilon\,  A_{d_{\sigma}}(\epsilon)
\left[f\left(\epsilon+\frac{\Phi}{2}\right)+f\left(\epsilon-\frac{\Phi}{2}\right)\right]\nonumber\\
&=\frac{1}{2\nu(0)\Omega}\sum_{\sigma k}  A_{d_{\sigma}}(\epsilon_k)
\left[f\left(\epsilon_k+\frac{\Phi}{2}\right)+f\left(\epsilon_k-\frac{\Phi}{2}\right)\right]
\end{align}
and the expression obtained in the effective equilibrium approach,
\be \label{nd-effeq}
n_d = \sum_\sigma \langle d_\sigma^\dag d_\sigma \rangle =\sum_{\sigma}{\cal G}_{d_{\sigma}{d}_{\sigma}^\dag}(\tau=0).
\ee
It is crucial to recall the definition of the Green's functions in the effective equilibrium approach (see Section \ref{sec:correspondence}). The Fourier representation of the (imaginary) time ordered Green's function is given by
\begin{align}
{\cal G}_{{\cal O}_{1}{\cal O}_{2}}(i\omega_{n})=\left\langle\left\{{\cal O}_{1},\frac{e^{i\omega_{n}0^{{+}}}}{i\omega_{n}-{\cal L}+{\cal L}_{Y}}{\cal O}_{2}\right\}\right\rangle,
\end{align}
and the retarded/advanced real time Green's function in frequency space is obtained by
\begin{align}\label{g-retadv}
G^{\text{ret/adv}}_{{\cal O}_{1}{\cal O}_{2}}(\omega)&=\left\langle\left\{{\cal O}_{1},\frac{1}{\omega\mp{\cal L}\pm i\eta}{\cal O}_{2}\right\}\right\rangle.
\end{align}

To prove the equivalence between Eqs.\ \eqref{nd-keldysh} and \eqref{nd-effeq}, we start with the Keldysh expression and evaluate the spectral function 
\be
A_{d_{\sigma}}(\epsilon)=-\frac{1}{\pi}\iim G^{\text{ret}}_{d_{\sigma}d_{\sigma}^\dag}(\epsilon) = \frac{1}{2\pi i}\bigg[ G^{\text{adv}}_{d_{\sigma}d_{\sigma}^\dag}(\epsilon)-G^{\text{ret}}_{d_{\sigma}d_{\sigma}^\dag}(\epsilon)\bigg].
\ee
With Eq.\ \eqref{g-retadv} we evaluate the retarded and advanced Green's functions involved, 
\begin{align}
G^{\text{ret}}_{d_{\sigma}d_{\sigma}^\dag}(\epsilon_k)
= \left\langle\left\{d
_{\sigma},\frac{1}{\epsilon_k-{\cal L}+i\eta}d^{\dag}_{\sigma}\right\}\right\rangle
= \frac{\sqrt{\Omega}}{t}
\langle\{d_{\sigma},\psi^{\dag}_{\alpha k\sigma}\}\rangle,
\end{align}
where, for simplicity, we are considering symmetric tunnel couplings and have set $t_1=t_{-1}=t$.
Similarly, one finds for the advanced Green's function
\begin{align}
G^{\text{adv}}_{d_{\sigma}d_{\sigma}^\dag}(\epsilon_k)
=\left\langle\left\{d
^{\dag}_{\sigma},\frac{1}{\epsilon_k-{\cal L}+i\eta}d_{\sigma}\right\}\right\rangle
=\frac{\sqrt{\Omega}}{t}\langle\{d^{\dag}_{\sigma},\psi_{\alpha k\sigma}\}\rangle. 
\end{align}
Utilizing these relations in the evaluation of the occupancy, we find
\begin{align}
&n_{d}=\\\notag
&\frac{1}{4\pi i}\frac{1}{t\nu(0)\sqrt{\Omega}}\sum_{\alpha k \sigma} \left(\langle\{d^{\dag}_{\sigma},\psi_{\alpha k\sigma}\}\rangle-\text{h.c.}\right)
f\left(\epsilon_k-\alpha\frac{\Phi}{2}\right).
\end{align}
Writing the Fermi function as a Matsubara sum, one finds
\begin{widetext}
\begin{align}
n_{d}&=\frac{1}{4\pi i}\frac{1}{t\nu(0)\sqrt{\Omega}}\frac{1}{\beta}\sum_{\alpha k \sigma \omega_{n}} \frac{e^{i\omega_{n}0^{{+}}}}{i\omega_{n}-\epsilon_k+\alpha\frac{\Phi}{2}}
\left(\langle\{d^{\dag}_{\sigma},\psi_{\alpha k\sigma}\}\rangle-\langle\{d_{\sigma},\psi^{\dag}_{\alpha k\sigma}\}\rangle\right)\notag\\
&=\frac{1}{4\pi i t\nu(0)\sqrt{\Omega}}\frac{1}{\beta}\sum_{\alpha k \sigma \omega_{n}} \bigg(\left\langle\left\{d^{\dag}_{\sigma},\frac{e^{i\omega_{n}0^{{+}}}}{i\omega_{n}+{\cal L}-{\cal L}_{Y}}\psi_{\alpha k\sigma}\right\}\right\rangle 
-\left\langle\left\{\frac{e^{i\omega_{n}0^{{+}}}}{i\omega_{n}-{\cal L}+{\cal L}_{Y}}\psi^{\dag}_{\alpha k\sigma},d_{\sigma}\right\}\right\rangle\bigg)\notag\\
&=\frac{1}{4\pi i t\nu(0)\sqrt{\Omega}}\frac{1}{\beta}\sum_{\alpha k \sigma \omega_{n}} \bigg(\left\langle\left\{d^{\dag}_{\sigma},\frac{e^{i\omega_{n}0^{{+}}}}{i\omega_{n}+{\cal L}-{\cal L}_{Y}}c_{\alpha k\sigma}\right\}\right\rangle + \left\langle\left\{d^{\dag}_{\sigma},\frac{e^{i\omega_{n}0^{{+}}}}{i\omega_{n}+{\cal L}-{\cal L}_{Y}}\frac{t}{\sqrt{\Omega}}\frac{1}{\epsilon_k+{\cal L}-i\eta}d_{\sigma}\right\}\right\rangle\notag\\
&\qquad\qquad -\left\langle\left\{\frac{e^{i\omega_{n}0^{{+}}}}{i\omega_{n}-{\cal L}+{\cal L}_{Y}}c^{\dag}_{\alpha k\sigma},d_{\sigma}\right\}\right\rangle
-\left\langle\left\{\frac{e^{i\omega_{n}0^{{+}}}}{i\omega_{n}-{\cal L}+{\cal L}_{Y}}\frac{t}{\sqrt{\Omega}}\frac{1}{\epsilon_k-{\cal L}+i\eta}d^{\dag}_{\sigma},d_{\sigma}\right\}\right\rangle\bigg).
\end{align}
The first and third terms can be combined, and recalling Eq.\ \eqref{eq:current1} they can be shown to cancel,
\begin{align}
&\frac{1}{4\pi i}\frac{1}{t\nu(0)\sqrt{\Omega}}\frac{1}{\beta} \sum_{\alpha k \sigma \omega_{n}}\left({\cal G}_{d_{\sigma}c_{\alpha k\sigma}^\dag}(i\omega_{n})-{\cal G}_{c_{\alpha k\sigma}d_{\sigma}^\dag}(i\omega_{n})\right)=-\frac{1}{2\pi et^{2}\nu(0)}\sum_{\alpha}\alpha I_{\alpha} =0,
\end{align}
since in the steady state we must satisfy $I_{1}=I_{-1}$. Therefore, this results in:
\begin{align}
n_{d}&=\frac{1}{4\pi i}\sum_{\alpha\omega_{n}\sigma}\frac{1}{\beta}\int_{-\infty}^{\infty}d\epsilon_{k}
\bigg(\left\langle\left\{d^{\dag}_{\sigma},\frac{e^{i\omega_{n}0^{{+}}}}{i\omega_{n}+{\cal L}-{\cal L}_{Y}}\frac{1}{\epsilon_k+{\cal L}-i\eta}d_{\sigma}\right\}\right\rangle
-\left\langle\left\{\frac{e^{i\omega_{n}0^{{+}}}}{i\omega_{n}-{\cal L}+{\cal L}_{Y}}\frac{1}{\epsilon_k-{\cal L}+i\eta}d^{\dag}_{\sigma},d_{\sigma}\right\}\right\rangle\bigg)\notag\\
&=\frac{1}{4\pi i}\sum_{\alpha\omega_{n}\sigma}\frac{1}{\beta}\int_{-\infty}^{\infty}d\epsilon_{k}
\bigg(\left\langle\left\{\frac{e^{i\omega_{n}0^{{+}}}}{i\omega_{n}-{\cal L}+{\cal L}_{Y}}d^{\dag}_{\sigma},\frac{1}{\epsilon_k+{\cal L}-i\eta}d_{\sigma}\right\}\right\rangle
-\left\langle\left\{\frac{e^{i\omega_{n}0^{{+}}}}{i\omega_{n}-{\cal L}+i\eta}d^{\dag}_{\sigma},\frac{1}{\epsilon_k+{\cal L}-{\cal L}_{Y}}d_{\sigma}\right\}\right\rangle\bigg).
\label{eq:intermediate}
\end{align}
\end{widetext}

In the next step, we will make use of the identity
\begin{align}\label{new-ident}
&\int_{-\infty}^{\infty} d\epsilon \frac{1}{{\epsilon}\pm{\cal L}\mp i\eta}\mathcal{O}=\pm i\pi \mathcal{O},
\end{align}
and we briefly outline its proof. It is convenient to use the spectral representation for this purpose and thus enumerate the set of eigenstates of $H$ by $\{|n\rangle\}$. With this, one obtains
\begin{align}
&\int_{-\infty}^{\infty}d\epsilon\, \langle m\vert\frac{1}{{\epsilon}\pm{\cal L}\mp i\eta}\mathcal{O}\vert n\rangle \notag\\
&=\int_{-\infty}^{\infty}d\epsilon\, \frac{1}{{\epsilon}\pm(\epsilon_{m}-\epsilon_{n})\mp i\eta}\langle m\vert\mathcal{O}\vert n\rangle d\epsilon.
\end{align}
After shifting the integration variable $\epsilon \rightarrow \epsilon \pm(\epsilon_{m}-\epsilon_{n})$, we can further simplify the above expression to yield
\begin{align}
&\int_{-\infty}^{\infty}d\epsilon \, \frac{1}{{\epsilon}\mp i\eta}\langle m\vert\mathcal{O}\vert n\rangle\\\nonumber
&=\left(\text{P.V.}\left[\int_{-\infty}^{\infty}\frac{d\epsilon}{\epsilon}\right]\pm i\pi\right)\langle m\vert\mathcal{O}\vert n\rangle.
\end{align}
Noting that the principal value integral vanishes concludes the proof of Eq.\ \eqref{new-ident}. With this identity in place, Eq.\ \eqref{eq:intermediate} can be brought into its final form
\begin{align}
n_{d}&=\frac{1}{4 }\sum_{\alpha\omega_{n}\sigma}\frac{1}{\beta}\bigg[\left\langle\left\{d^{\dag}_{\sigma},\frac{e^{i\omega_{n}0^{{+}}}}{i\omega_{n}+{\cal L}-{\cal L}_{Y}}d_{\sigma}\right\}\right\rangle\notag\\
&\qquad +\left\langle\left\{\frac{e^{i\omega_{n}0^{{+}}}}{i\omega_{n}-{\cal L}+{\cal L}_{Y}}d^{\dag}_{\sigma},d_{\sigma}\right\}\right\rangle\bigg]\notag\\
&=\frac{1}{\beta}\sum_{\omega_{n}\sigma}{\cal G}_{d_{\sigma}d_{\sigma}^\dag}(i\omega_n)=\sum_{\sigma}{\cal G}_{d_{\sigma}{d}_{\sigma}^\dag}(\tau=0).
\end{align}
This demonstrates that the Schwinger-Keldysh approach and the effective equilibrium formulation reproduce the same expression for the occupancy on the quantum dot. We note that the general expressions (53) and (B7) are particularly useful since they
can be directly applied through numerical procedures such as numerical RG.\cite{Anders_2008,Anders_2010}

\section{Derivation of $n_{d}$ using the Keldysh approach}
\label{app:3} 
The basic equation which will constitute the starting point of our proof is given by Eq.(15) in Ref.\ \onlinecite{Meir_Wingreen}. The current in lead $\alpha=\pm1$ for spin projection $\sigma$ is given by
\begin{align}
I_{\al
\sigma}(t)=&-\alpha\frac{2e}{\hbar}\int_{-t_{0}}^{t}dt_{1}\int\frac{d\ep}{2\pi}\text{Im}\bigg\{e^{-i\ep(t_{1}-t)}\Gamma^{\al}(t,t_{1})\notag\\
&\times\left[G^{<}_{\sigma\sigma}(t,t_{1})+f_{\al}(\ep)G^{r}_{\sigma\sigma}(t,t_{1})\right]\bigg\},
\end{align}
where we have restricted ourselves to a single level on the dot having a spin index $\sigma$.
The occupation number($n_{d_{\sigma}}$) of the dot in the state $\sigma$ obeys the following differential equation:
\begin{align}
\frac{dn_{d_{\sigma}}(t)}{dt}=\frac{1}{-e}\sum_{\al} \al I_{\al\sigma}.
\label{eq:number}
\end{align}
In steady state ({\it i.e.}, t=0) we get 
\begin{align}
\frac{dn_{d_{\sigma}}(t)}{dt}\bigg|_{t=0}=0.
\end{align}
This implies 
\begin{align}
\sum_{\alpha}\int_{-\infty}^{0}dt_{1}\int\frac{d\ep}{2\pi}&\text{Im}\bigg\{e^{-i\ep t_{1}}\Gamma^{\al}(0,t_{1})\notag\\
&\left[G^{<}_{\sigma\sigma}(0,t_{1})+f_{\al}(\ep)G^{r}_{\sigma\sigma}(0,t_{1})\right]\bigg\}=0.
\label{eq:derivative}
\end{align}
Since in steady state there exists time translational invariance, we have $G^{<}_{\sigma\sigma}(0,t_{1})=G^{<}_{\sigma\sigma}(0-t_{1})$ and $G^{r}_{\sigma\sigma}(0,t_{1})=G^{r}_{\sigma\sigma}(0-t_{1})$. Further we can set $\eta\rightarrow 0$ and $t_{0}\rightarrow -\infty$, in the sense defined by the open system limit, right from the beginning. This in turn implies that $\Gamma^{\al}(0,t_{1})=\Gamma^{\al}=\pi\nu(0)t^{2}=\Gamma/2$. Let us focus on the first term in the expression above
\begin{align}
{\cal A}=&\sum_{\alpha}\int_{-\infty}^{0}dt_{1}\int\frac{d\ep}{2\pi}\text{Im}\bigg\{e^{-i\ep t_{1}}\Gamma^{\al}G^{<}_{\sigma\sigma}(-t_{1})\bigg\}\notag\\
&=-\frac{i}{2}\sum_{\alpha}\Gamma^{\al}\int_{-\infty}^{0}\int\frac{d\ep}{2\pi}\bigg\{e^{-i\ep t_{1}}G^{<}_{\sigma\sigma}(-t_{1})\notag\\
&\qquad\qquad-e^{i\ep t_{1}}\left[G^{<}_{\sigma\sigma}(-t_{1})\right]^{\ast}\bigg\}\notag\\
\intertext{Using the fact $\left[G^{<}_{\sigma\sigma}(-t_{1})\right]^{\ast}=-G^{<}_{\sigma\sigma}(t_{1})$}
&{\cal A}=-\frac{i}{2}\sum_{\alpha}\Gamma^{\alpha}\int\frac{d\ep}{2\pi}G^{<}_{\sigma\sigma}(\ep)\notag\\
&=-\frac{i}{2}\Gamma G^{<}_{\sigma\sigma}(0)\notag\\
&=\frac{1}{2}\Gamma \left\langle d^{\dag}(0)d(0)\right\rangle=\frac{1}{2}\Gamma n_{d_{\sigma}}.
\end{align}
Similarly for the second  term we get
\begin{align}
{\cal B}=&\sum_{\alpha}\int_{-\infty}^{0}dt_{1}\int\frac{d\ep}{2\pi}\text{Im}\bigg\{e^{-i\ep t_{1}}f_{\al}(\ep)\Gamma^{\al}G^{r}_{\sigma\sigma}(-t_{1})\bigg\}\notag\\
&=\frac{\Gamma}{2}\sum_{\alpha}\int\frac{d\ep}{2\pi}f_{\al}(\ep)\text{Im}\bigg\{\int_{-\infty}^{0}dt_{1}e^{i\ep t_{1}}G^{r}_{\sigma\sigma}(-t_{1})\bigg\}\notag\\
&=\frac{\Gamma}{2}\sum_{\alpha}\int\frac{d\ep}{2\pi}f_{\al}(\ep)\text{Im}\bigg\{G^{r}_{\sigma\sigma}(\ep)\bigg\}\notag\\
&=\Gamma\int\frac{d\ep}{2}\underbrace{\sum_{\alpha}\frac{f_{\al}(\ep)}{2}}_{f^{\text{eff}}(\ep,\Phi)}\underbrace{\frac{1}{\pi}\text{Im}\bigg\{G^{r}_{\sigma\sigma}(\ep)\bigg\}}_{-A_{d_{\sigma}}(\ep)}\notag\\
&=-\frac{1}{2}\Gamma\int d\ep f^{\text{eff}}(\ep,\Phi) A_{d_{\sigma}}(\ep).
\end{align}
 Combining the two terms and putting in Eq.\ (\ref{eq:derivative}) we get
\begin{align}
n_{d_{\sigma}}=\int d\ep f^{\text{eff}}(\ep,\Phi) A_{d_{\sigma}}(\ep),
\end{align}
which implies that
\begin{align}
n_{d}=\sum_{\sigma}n_{d_{\sigma}}&=\int d\ep f^{\text{eff}}(\ep,\Phi) \sum_{\sigma}A_{d_{\sigma}}(\ep)\notag\\
&=\int d\ep f^{\text{eff}}(\ep,\Phi) A_{d}(\ep).
\end{align}
This is identical to the expression obtained via the spectral representation in Appendix \ref{app:2}.

\end{document}